\documentclass[preprint2]{aastex}

\usepackage{lscape}

\newcommand{\beq}{\begin{equation}}
\newcommand{\eeq}{\end{equation}}
\newcommand{\chandra}{{\it Chandra }}

\shorttitle{Prevalence of X-ray variability in the CDFS}
\shortauthors{Paolillo et al.}

\begin{document}


\title{Prevalence of X-ray variability in the {\it Chandra} Deep Field South}


\author{M. Paolillo\altaffilmark{1,2}, E. J. Schreier\altaffilmark{1,3} , R. Giacconi\altaffilmark{3,4}, A. M. Koekemoer\altaffilmark{1}, N. A. Grogin\altaffilmark{4}}
\affil{}

\email{}


\altaffiltext{1}{Space Telescope Science Institute, Science Division, 3700 San Martin Drive,  Baltimore, MD 21218, USA}
\altaffiltext{2}{Universit\`a degli Studi Federico II, Dip.di Fisica, C. U. Monte S.Angelo, via Cintia, 80126 Napoli, ITALY}
\altaffiltext{3}{Associated Universities Inc., 1400 16$^{th}$ Street N.W., Washington DC 20036, USA}
\altaffiltext{4}{Johns Hopkins University, Department of Physics and Astronomy, 3701 San Martin Drive, Baltimore, MD 21218, USA}


\begin{abstract}
We studied the X-ray variability of sources detected in the {\it Chandra} Deep Field South \citep{G02}, nearly all of which are low to moderate $z$ AGN \citep{Tozzi01}. We find that 45\% of the sources with $>100$ counts exhibit significant variability on timescales ranging from a day up to a year. The fraction of sources found to be variable increases with observed flux, suggesting that $>90\%$ of all AGNs possess intrinsic variability. 
We also find that the fraction of variable sources appears to decrease with increasing intrinsic absorption; a lack of variability in hard, absorbed AGNs could be due to an increased contribution of reflected X-rays to the total flux.
We do not detect significant {\it spectral} variability ($\Delta \Gamma>0.2$) in the majority ($\sim 70\%$) of our sources. In half of the remaining $30\%$, the hardness ratio is anti-correlated with flux, mimicking the high/soft--low/hard states of galactic sources.
The X-ray variability appears anti-correlated with the luminosity of the sources, in agreement with previous studies. High redshift sources, however, have larger variability amplitudes than expected from extrapolations of their low-z counterparts, suggesting a possible evolution in the accretion rate and/or size of the X-ray emitting region. Finally, we discuss some effects that may produce the observed decrease in the fraction of variable sources from $z=0.5$ out to $z=2$.
 
\end{abstract}




\section{Introduction}
Active Galactic Nuclei (AGN) are found to vary on timescales ranging from minutes to years. Variability studies have proven to be an important tool to investigate the innermost regions of AGNs, which cannot be resolved with the current generation of astronomical instruments. The timescale of the variability, its spectral dependence and the correlation between different line and continuum components provide fundamental clues on the nature of the physical processes which occur near to the central Black Hole (BH) and on the size and relation between the different regions producing the observed emission \cite[see][for a comprehensive review]{Ulrich97}.
 
In the X-ray band, AGNs show faster variability than in any other band, consistent with the X-ray emission originating from a small region very close to the central BH \citep[see][]{Mushotzky93}. Several promising models have been proposed to explain the fast intrinsic variability and its characteristics: a single coherent oscillator \cite[e.g.][]{Almaini00}, variable decay shot-noise due to a superposition of individual flares \cite[e.g.][]{Lehto89}, and bright rotating spots spiraling around the BH \citep{Abramowicz91}. All these models predict a dependence on the BH mass, accretion rate and the size of the X-ray emitting region which may in turn explain the observed dependence of variability on the AGN luminosity \citep[e.g.][]{Law93,GML93,Nandra97} and spectral properties \citep{Nandra97,Turner99,Fio98}.
 
Most variability studies in the X-ray band are based on samples of bright and nearby AGNs for which good quality data are available from past X-ray missions. Moreover they usually sample short timescales from hours to days. Only for the nearest sources have long monitoring campaigns allowed a detailed study of long-term variability up to several years. The {\it Chandra} Deep Field South \cite[hereafter CDFS]{G01} represents one of the deepest observations of the Universe in the X-ray band (the only deeper sample is in the Northern {\it Chandra} Deep Field). \cite{Rosati02} detected 346 sources in the CDFS, resolving $>90\%$ of the hard X-ray background. 
The X-ray luminosities and spectral properties of the sources revealed that $\sim 80\%$ of them are AGNs \citep{G01,Tozzi01}.
Optical follow-up studies showed that they are equally divided among moderate redshift ($z<1.6$) Type II AGNs with large intrinsic absorption ($\log N_H>21.5$ cm$^{-2}$), and unabsorbed Type I AGNs covering a large range of redshifts up to $z\sim 3.6$, both hosted by a broad range of galaxy types \citep{Schreier01,Koekemoer02,Rosati02}. The remaining $20\%$ of the sample is composed of a mixture of normal galaxies, starburst galaxies, galaxy clusters \citep{G01,Koekemoer02} and individual off-nuclear X-ray sources \citep{Horn04}.

Being composed of several concentric exposures obtained over a period of $\sim 1$ year, the CDFS represent an ideal dataset to study the X-ray variability of faint and distant AGNs, on timescales ranging from a day up to several months. In the current study we measure the variability properties of the AGN sample and correlate the variability with other X-ray properties (flux, luminosity, hardness ratio, etc.), compare the properties of these moderate-$z$ AGN to those of the well studied nearby AGN population and study their evolution with look-back time.
The paper is organized as follows: in $\S$ 2 we describe the dataset, in $\S$ 3 we illustrate the variability analysis, in $\S$ 4 and 5 we present and discuss our results and in $\S$ 6 we draw the conclusions.

Throughout the paper we assume a cosmological model with $H_0=70$ km s$^{-1}$ Mpc$^{-1}$, $\Omega_m=0.3$ and $\Omega_\Lambda=0.7$.

\section{The data}
The CDFS \citep{G01} was observed 11 times by the Chandra X-ray Observatory from October 1999 to December 2000. The diary of the observations and their properties are described in detail in \cite{G02}, \cite{Tozzi01}, \cite{G01}; we summarize in Table \ref{obstab} the properties of each exposure that are relevant to this paper. Each observation was filtered to include only standard ASCA event grades, hot pixels and columns were removed and high background intervals were subtracted; all exposures were then added to produce a "stacked observation" of 942 ks \citep[see][for details]{G02}.
Exposure maps were produced for each individual observation in both the soft (0.5-2.0 keV) and hard (2.0-7.0 keV) bands, and then combined using the appropriate shift and weighting for individual exposure times. Throughout this paper we used the hard band exposure maps; we will discuss the uncertainties introduced by this choice in $\S$\ref{uncertainty}.
The stacked observation was used to detect X-ray sources both through a modified version of the SExtractor algorithm \citep{Bertin96} and the WAVDETECT  \citep{dobrzycki1999,freeman2001} algorithm included in the CIAO\footnote{http://cxc.harvard.edu/ciao} software. The final catalogues, including 346 sources, are presented in \cite{G02}.

\section{Data analysis}
\subsection{Lightcurve extraction}
\label{ltc_sec}
To study the temporal variability of sources detected in the CDFS we extracted counts from both the 11 individual exposures and the stacked observation for each source in the \cite{G02} catalogue, in the 0.5-7.0 keV band. We adopted the 95\% encircled-energy extraction radii $R_S$ used by \cite{G02}, which take into account the broadening of the PSF full width half maximum (FWHM) with off-axis angle. Similarly the local background was measured in an annulus with inner and outer radii of $R_S+2$ and $R_S+12$ pixels respectively, after masking out overlapping sources. The source counts were then background-subtracted and exposure corrected to match the observing condition at the ACIS-I aimpoint. Errors were computed using the Gehrels approximation $\sigma\approx 1+\sqrt{n + 0.75}$ \citep{Gehrels86}, which provides a more reliable estimate in the low count regime.

\begin{figure*}
\epsscale{2.0}
\plotone{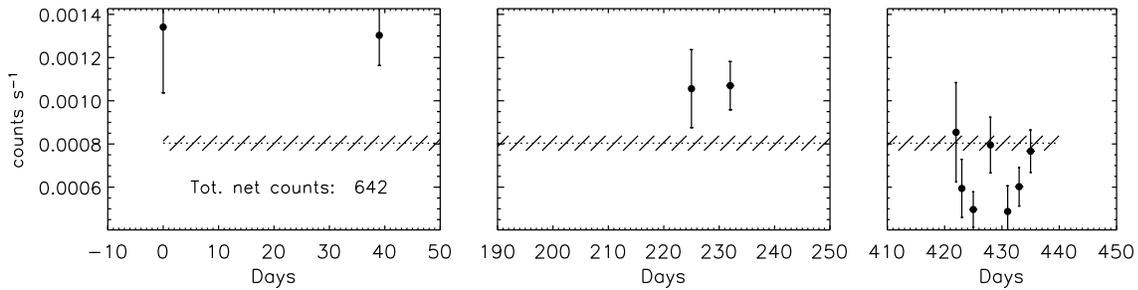}
\caption{ A typical lightcurve of a variable CDFS source with $\sim 600$ counts. Each bin
represent an individual Chandra observation. The dotted line represents the average count
rate, while the shaded region shows the error on the average value. The gaps between panels represent extended periods without observations. \label{ltc}}
\end{figure*}

Even though the observations which compose the CDFS have similar aimpoints, they were designed to have small shifts between exposures, to allow a more uniform coverage of the gaps between the ACIS CCDs; furthermore they have different roll angles. For this reasons (and because of the square shape of the ACIS-I detector) sources  near the edge of the CDFS field of view (FOV) may fall inside the CCDs in some observations, while lying outside the detector or partially on the edge in others. In order to maximize the number of data bins for each source, we decided to retain those bins corresponding to observations in which: 1) less than 10\% of the source area and 2) less than 50\% of the background region fell outside the FOV. These cases however are a minority (15\% of all sources) and we verified that they do not affect our analysis, as discussed in $\S$ \ref{uncertainty}. We excluded from the subsequent analysis 6 sources which had only one acceptable observation. 
An example of a typical lightcurve obtained in this way is shown in Figure \ref{ltc}.

\subsection{Variability estimates}
\label{var_est}
A commonly used quantity to estimate variability in literature is the lightcurve $\chi^2$ defined as:
\beq
\label{chi2}
\chi^2={1\over N_{obs}-1}\sum_{i=1}^{N_{obs}}\frac{(n_i-\overline{n})^2}{\sigma_i^2}
\eeq
where $N_{obs}$ is the number of observations in which the sources falls inside the FOV, $n_i$ and $\sigma_i$ are the number of counts and its error measured in the $i^{th}$ observation after background subtraction and exposure correction, and $\overline{n}$ is the average number of counts extracted from the stacked observation.

\begin{figure*}
\epsscale{1.0}
\plotone{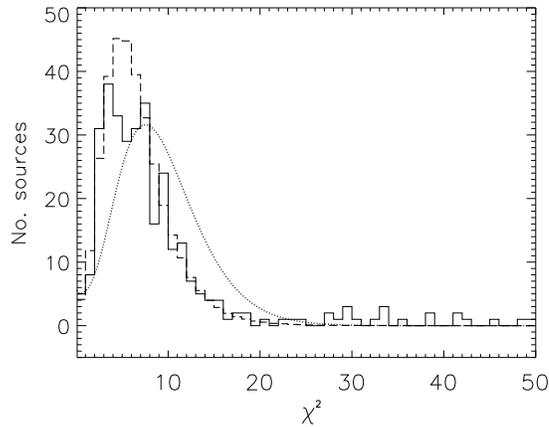}
\caption{ $\chi^2$ distribution of the CDFS source lightcurves (continuous line) compared to those derived from simulations (dashed line) and to the one expected in the case of a
normal distribution (dotted line). The distributions are normalized to the total number of sources. Note that even though the plot is truncated at $\chi^2=50$, there are several sources with larger $\chi^2$.\label{chisq}}
\end{figure*}

A large part of the sources detected in the CDFS, however, have less than 100 counts, which implies on average $\leq 10$ counts/observation. It is well known that, in such low counts regime, the Poissonian distribution of X-ray events significantly differs from a Gaussian distribution. This difference is further enhanced by our choice to use the Gehrels error formula discussed above. The final counts (background subtracted and exposure corrected), obtained through the standard error propagation, are not normally distributed anymore so that the probability estimates derived from the $\chi^2$ distribution in such approximation are not accurate (a detailed discussion of this issue can be found in the IRAF/PROS Xray package\footnote{http://iraf.noao.edu/scripts/irafhelp?explain\_errors}, also see \citealp{Nousek89,Wheaton95}).
To properly deal with the small number of counts, we performed 1000 Monte-Carlo simulated observations of each source, mimicking the exact conditions of each exposure. The simulated source is composed of the superposition of source and background counts extracted from a Poisson distribution with the same median as the real source, in the assumption that the flux is non-variable. The details of the simulations are described in Appendix \ref{app_simul}. The simulated counts were background subtracted, exposure corrected and divided by the exposure times exactly as done for the real sources. Using the same Gehrels error approximation used for real observations, we thus derived 1000 simulated $\chi^2$ for each source.
In Figure \ref{chisq} we compare the $\chi^2$ distribution of the CDFS sources with those obtained from our simulation and with the one expected in the Gaussian approximation (in which case each source $\chi^2$ would follow a probability distribution given by an incomplete gamma function $\Gamma(\nu/2,\chi^2/2)$ where $\nu$ are the degrees of freedom; \citealp[see][]{numrec}).
The actual $\chi^2$ distribution has a lower median than the Gaussian prediction. This is expected since the Gehrels error approximation is properly an upper limit and always larger than the Gaussian estimate $\sqrt{n}$, resulting in lower $\chi^2$.
On the other end the simulated distribution agrees well with the observed one, except for $\chi^2>20$ where the presence of variable sources (not contemplated in the simulations) produces an extended tail; note that the larger peak value in the simulated distribution is due to the fact that the histograms are normalized to the total number of sources.

The variability of the sources was estimated comparing the $\chi^2$ measured from the lightcurve to those obtained from the simulated observations. We derived two variability estimates: in the first one the comparison is made between the total $\chi^2$, i.e. summing the deviations in all bins as in expression (\ref{chi2}). The second estimate is obtained using just the bin in which we observe the maximum deviation from the mean, i.e. using the quantity 
\beq
\chi^2_{max}=\max_{i\in1,\cdots,N_{obs}}\left\{\frac{(n_i-\overline{n})^2}{\sigma_i^2}\right\}
\eeq
While the first estimate provides an average estimate over the whole lightcurve, and thus is better suited for sources which have small and repeated fluctuations, the second one allows to detect variability in sources exhibiting rare bursts, which may be undetected when we average over the whole lightcurve. We classify as variable those sources whose probability to obtain a $\chi^2$ smaller than the measured one, in the assumption of a constant count rate, is $P(\chi^2)>95\%$ for the averaged estimate or $P_{max}(\chi^2)>99\%$ in the maximum deviation estimate. A higher limit was adopted for $P_{max}(\chi^2)$ since its distribution is skewed toward larger values by definition; furthermore $P_{max}(\chi^2)$ may be more affected by deviations occurring in a single observation than $P(\chi^2)$, which is a more robust estimate since it is averaged on the whole lightcurve. 
This choice however does not significantly affect our results: the use of $P(\chi^2)>99\%$ would change the variable/non-variable classification of only 4 sources (2 if we consider objects with $>100$ counts, as done in most of our analysis).

\begin{figure*}
\epsscale{1.0}
\plotone{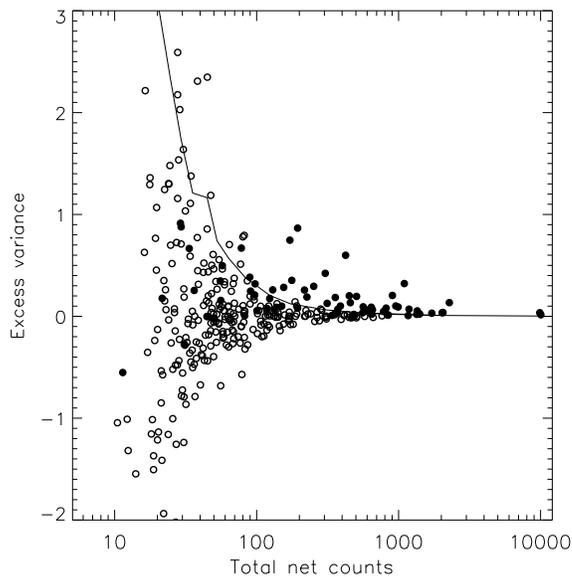}
\caption{Excess variance vs total net counts. Empty (filled) circles represent non-variable (variable) sources. The continuous line represents the 95\% upper limit on the excess variance due to statistical uncertainties for a constant source, as derived from simulations. \label{var_vs_cnts}}
\end{figure*}

To obtain a quantitative measure of the variability amplitude we calculated the ``excess variance'' of the CDFS sources \citep{Nandra97,Turner99}:
\beq
\label{excess_var}
\sigma^2_{rms}={{1}\over{N_{obs}\overline{n}^2}}\sum_{i=1}^{N_{obs}} [(n_i-\overline{n})^2-\sigma^2_i]
\eeq
$\sigma_{rms}$ measures the fraction of the total flux per bin which is variable, after subtracting the statistical error. We also calculated a ``maximum excess variance'' for each source, defined as:
\beq
\sigma^2_{max}=\max_{i\in1,\cdots,N_{obs}}\left\{{(n_i-\overline{n})^2-\sigma^2_i}\over{\overline{n}^2}\right\}
\eeq
The accuracy of $\sigma_{rms}$ as a variability estimator depends on the S/N ratio of the sources, as shown in Figure \ref{var_vs_cnts}. Faint sources with $\lesssim 100$ net counts lie mostly below the $95\%$ upper limit derived from simulations (continuous line) indicating that statistical uncertainties dominate the excess variance estimates, thus making $\sigma_{rms}$ unreliable. At higher count levels our sensitivity improves and an increasingly larger fraction of the sources is detected as variable. 
In the high statistics regime $\sigma_{rms}$ is well correlated with the variability estimates obtained from the $\chi^2$ analysis, in the sense that all sources with $\sigma_{rms}$ above the 95\% limit are found to be variable. Below 100 counts such correlation does not hold anymore: this is due to the fact that while the $\chi^2$ analysis correctly estimates the probability that a source is variable through comparison with the simulations, the correction for the statistical error in expression (\ref{excess_var}) is based on a Gaussian approximation which fails in the low counts regime. Also note that the $95\%$ limit shown in Figure \ref{var_vs_cnts} is an {\it average} value: the actual limit for each individual source depends on the observing conditions (exposure time, position within the FOV, PSF etc.) which may vary, thus explaining why a few variable sources with $>100$ counts are scattered below this limit.

\cite{Almaini00} point out that expression (\ref{excess_var}) represents the best variability estimator only for identical measurements errors ($\sigma_i=$constant) and otherwise a numerical approach should be used. However, in our case, their method is not strictly correct either because our measurements are not normally distributed. We thus decided to adopt the simpler analytical approach; the errors in our variability estimates are dominated anyway by statistical uncertainties due to the low count rate, as discussed in next section.

\subsection{Sources of uncertainty} 
\label{uncertainty}
It is possible for the intrinsic variability present in our sources to be masked or enhanced by statistical errors or systematic effects. In Appendix \ref{uncertainty_app} we discuss the principal sources of systematic uncertainty which may affect CDFS sources: QE degradation, exposure corrections, PSF variations, corrections for sources on edges. We find that these effects influence our count rate estimates by less than 10\%.  This implies that the main source of uncertainty in the majority of our sources is the statistical error. In fact, statistical errors $\sigma_i\simeq \sqrt{n_i}$ drop below 10\% only for sources with $n_i>100$ counts per observation on average, i.e. $n_{tot}>1100$ total counts. The CDFS contains just 11 such sources and all of them exhibit much larger flux variations. Note however that, as pointed out in the previous section, we can draw significant conclusion for all sources with $>100$ total counts. In the discussion below we consider the effect of these uncertainties on our results.

\section{Results}
\subsection{Variable fraction}
\label{varfrac_sec}
Through the $\chi^2$ analysis we are able to detect variability in 74 out of 340 sources, i.e. $\sim 20\%$ of the sample, in the 0.5-7 keV band. However our ability to detect flux variations depends on the S/N ratio of the sources: for sources with $<100$ counts the observed flux variations are dominated by statistical fluctuations ($\S$ \ref{var_est}). In the following analysis we therefore concentrate on sources with $>100$ total counts, where the statistical uncertainties drop below $\sigma_{rms}\sim 50\%$ (continuous line in Figure \ref{var_vs_cnts}). The faintest flux variations that we detect in these sources corresponds to $\sigma_{rms}\sim 10\%$.

To better restrict our variability analysis to the AGN population, we can also exclude from the sample sources which are not a-priori expected to possess any temporal variability. \cite{G02} found that 19 of the CDFS sources were spatially extended and thus are unlikely to be variable. Furthermore at low luminosities ($L_X\lesssim 10^{42}$ erg s$^{-1}$) the X-ray population is believed to contain a large fraction of starburst and/or normal galaxies \citep[e.g.][]{Alexander02}. We derived intrinsic X-ray luminosities in the 0.5-7 keV band using the photometric redshifts calculated by \cite{Mobasher03} using Hubble-ACS and ground-based optical and NIR data acquired as part of the GOODS collaboration \citep{Giavalisco03}. Luminosities were k-corrected assuming a power law spectrum with $\Gamma =1.8$\footnote{The use of more complex models including an average absorption to compute the k-correction does not significantly change the results.} \citep{Alexander03}. 

\begin{figure*}
\epsscale{1.0}
\plotone{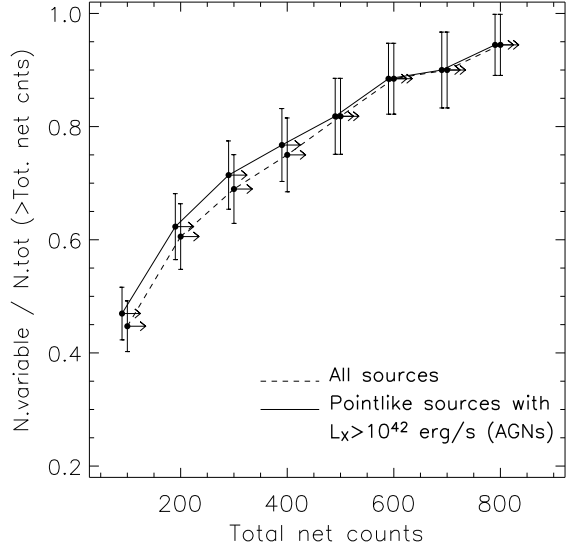}
\caption{Fraction of variable sources vs net counts. The arrows show increase in the fraction of sources which are found to be variable in the 0.5-7 keV band, selecting increasingly higher S/N subsamples (i.e. sources with total counts$>Net~counts$).
The upper curve is slightly shifted for visualization purposes.\label{varfrac}}
\end{figure*}

The dependence of the fraction of variable sources on the total counts is shown in Figure \ref{varfrac}. 
The fraction increases from 45\% considering sources with $> 100$ counts, to 94\% for sources with $> 800$ counts. Excluding extended sources and those with L$_X<10^{42}$ erg s$^{-1}$ (dashed line) the result does not change substantially, confirming that AGNs dominate our sample. This increase in the variable fraction with counts indicates that, as we become more sensitive to detecting variability, we systematically find a larger number of variable sources. Even though the limited number of objects requires caution in extrapolating the variable fraction to higher counts, Figure \ref{var_vs_cnts} further suggests that all sources with $>1000$ counts may be variable.

\subsection{Variability dependence on hardness ratio}
\label{HRvar_sec}
\begin{figure*}
\epsscale{1.2}
\plotone{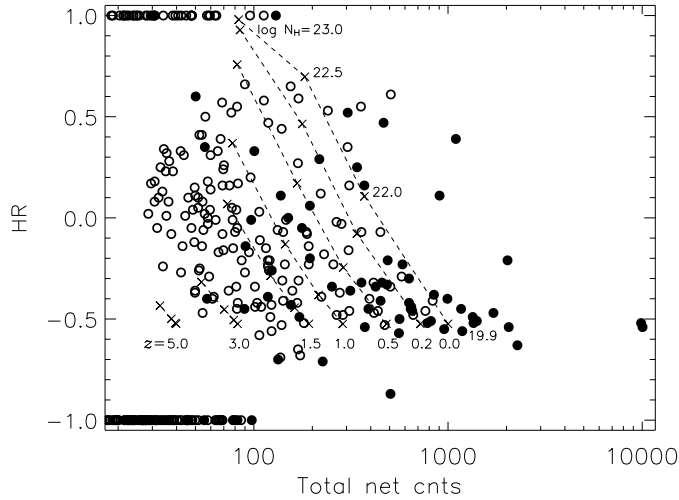}
\caption{Hardness ratio vs net counts. The filled and empty circles represent variable and non-variable sources respectively. Each dotted line represent the track followed by a source at a given redshift ($z=0,0.2,0.5,1.0,1.5,3,5$) as the intrinsic absorption  increases ($\log N_H=19.9,22,22.5,23$ cm$^{-2}$). The tracks are normalized to $f_X\simeq 10^{-14}$ erg s$^{-1}$ cm$^{-2}$ (1000 counts) at $z=0$; fainter normalization would shift the tracks to the left.
\label{HR_vs_cnts}}
\end{figure*}

The increase in the fraction of variable objects with total flux, discussed in the previous section, is obviously partly due to the increased S/N ratio of the sources. A second effect, however, seems to be present. Comparing the hardness ratio HR=H-S/H+S, where H and S are the counts measured from the stacked observation in the soft (0.5-2 keV) and hard (2-7 keV) bands, with the net source counts (Figure \ref{HR_vs_cnts}), we see that the average HR decreases as the sources grow stronger. Furthermore the majority of variable objects are soft (HR$\lesssim -0.2$). 
Thus, choosing higher flux sources systematically selects a larger fraction of the soft population, which appears to contain the majority of variable sources. 

The counts vs HR distribution can be explained by the presence of absorbing material near the central AGN. In Figure \ref{HR_vs_cnts} we plot the tracks followed by a source at given redshift as we increase the intrinsic absorption. We assumed a source with power law spectrum of $\Gamma =1.8$ normalized to 1000 counts ($f_X\simeq 10^{-14}$ erg s$^{-1}$ cm$^{-2}$) at $z=0$ and a galactic column density of $N_H=8\times 10^{19}$ cm$^{-2}$. The effect of adding obscuring material is to make the source fainter and harder, to an extent which depends on the redshift of the source. The plot shows that the bulk of the variable population is characterized by lower column densities than the non-variable one.

\begin{figure*}
\epsscale{1.0}
\plotone{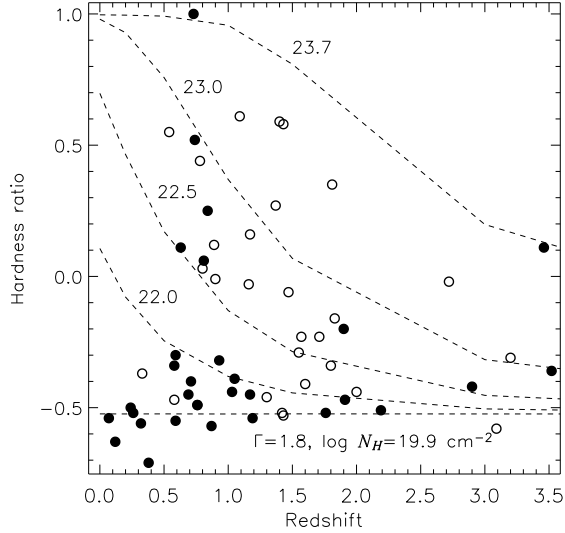}
\caption{Hardness ratio vs redshift for variable (filled circles) and non-variable (empty circles)
sources. The dashed lines show the HR expected for a source with $\Gamma=1.8$ and increasing 
intrinsic absorbing column ($\log N_H=22,22.5,23,23.7$ cm$^{-2}$). The lower level (horizontal line) corresponds to galactic absorption of $N_H=8\times 10^{19}$ cm$^{-2}$.\label{abs}}
\end{figure*}

Considering the subsample with available photometric redshifts the connection between variability and absorption is even clearer. In Figure \ref{abs} we plot the HR vs redshift of sources with $> 100$ counts, excluding the extended sources and those with $L_X<10\times 10^{42}$ erg s$^{-1}$ (see $\S$ \ref{varfrac_sec}). Variable sources appear systematically softer than non-variable ones at all redshifts. Moreover, comparing the HR of the sources with the one expected assuming an intrinsic power law spectrum ($\Gamma=1.8$) and increasing levels of photoelectric absorption, we find that 2/3 of variable sources have $N_H<10^{22.5}$ cm$^{-2}$. The opposite effect holds for non-variable sources.

We tried different approaches to determine if the lower fraction of variable sources in the {\it absorbed} ($\log N_H> 22.5$ cm$^{-2}$) subsample is an intrinsic feature of this population or is a selection effect due to the fact that such sources are also fainter than the corresponding {\it unabsorbed} ($\log N_H< 22.5$ cm$^{-2}$) counterparts and thus less likely to be detected as variable.

First, to measure the effect that a lower count-rate has on the fraction of variable sources that we detect, we artificially rescaled the counts measured in each exposure for the {\it unabsorbed} population, so that their median counts match the median counts of the {\it absorbed} subsample. We note that a simple rescaling by a constant factor of the measured counts would not change the S/N ratio of the sources, and thus would not correctly simulate a fainter population. In order to correctly model the count distribution, we re-extracted the source counts from a Poisson distribution whose mean is the rescaled value. We then re-estimate their variability through the procedure outlined in $\S$ \ref{var_est}. This approach preserves the intrinsic variability of the sources while simulating the effects of a lower statistic. In the assumption that {\it absorbed} and {\it unabsorbed} sources have the same intrinsic variability and that the observed differences are only due to a different average S/N ratio, we would expect the rescaled {\it unabsorbed} population to have the same variable fraction as the {\it absorbed} one. We find a variable fraction of 45\% in the rescaled {\it unabsorbed} population. Assuming that this is the fraction of variable sources expected in the {\it absorbed} population due only to the lower S/N ratio, we calculate that the probability of finding 9 variable sources out of 29 with $>100$ counts, as observed in the {\it absorbed} population, is $<5\%$.

\begin{figure*}
\epsscale{1.0}
\plotone{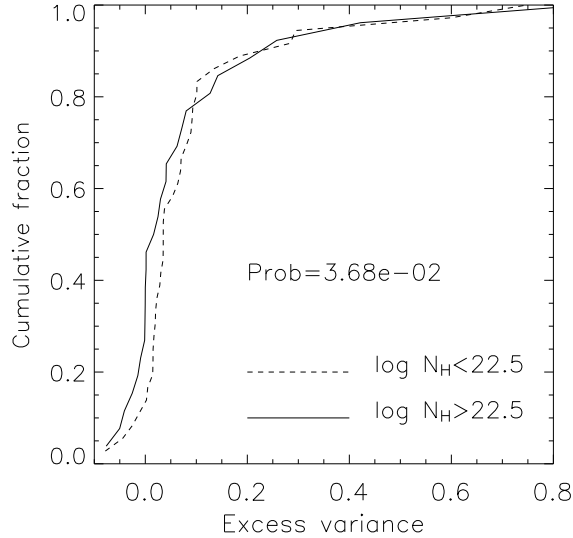}
\caption{Cumulative distribution of {\it absorbed} ($\log N_H> 22.5$ cm$^{-2}$) and {\it unabsorbed} ($\log N_H< 22.5$ cm$^{-2}$) sources vs excess variance, showing the probability that the two subsamples are extracted from the same population according to a KS test.\label{excess_var_KS}}
\end{figure*}

Second, we compared the excess variance distribution of {\it absorbed} and {\it unabsorbed} sources. Being corrected for statistical errors, the excess variance is less affected by the count-rate bias discussed before. Furthermore, since $\sigma^2_{rms}$ represents the fractional variation of the total flux, it is not expected to change due only to photoelectric absorption  because both the average flux and its fluctuations are attenuated by the same factor. In Figure \ref{excess_var_KS} we show the cumulative $\sigma^2_{rms}$ distributions for the {\it absorbed} and {\it unabsorbed} subsamples. The null hypothesis that the two subsamples are extracted from the same distribution is rejected at the 96\% level.  


While none of these tests is conclusive, they suggest that the lower variability observed in the hard absorbed population is an intrinsic feature and is not due to selection effects.

\subsection{Characterizing the variability}
\label{char_sec}
\subsubsection{Total flux variability}
\begin{figure*}[!h]
\epsscale{2.}
\plottwo{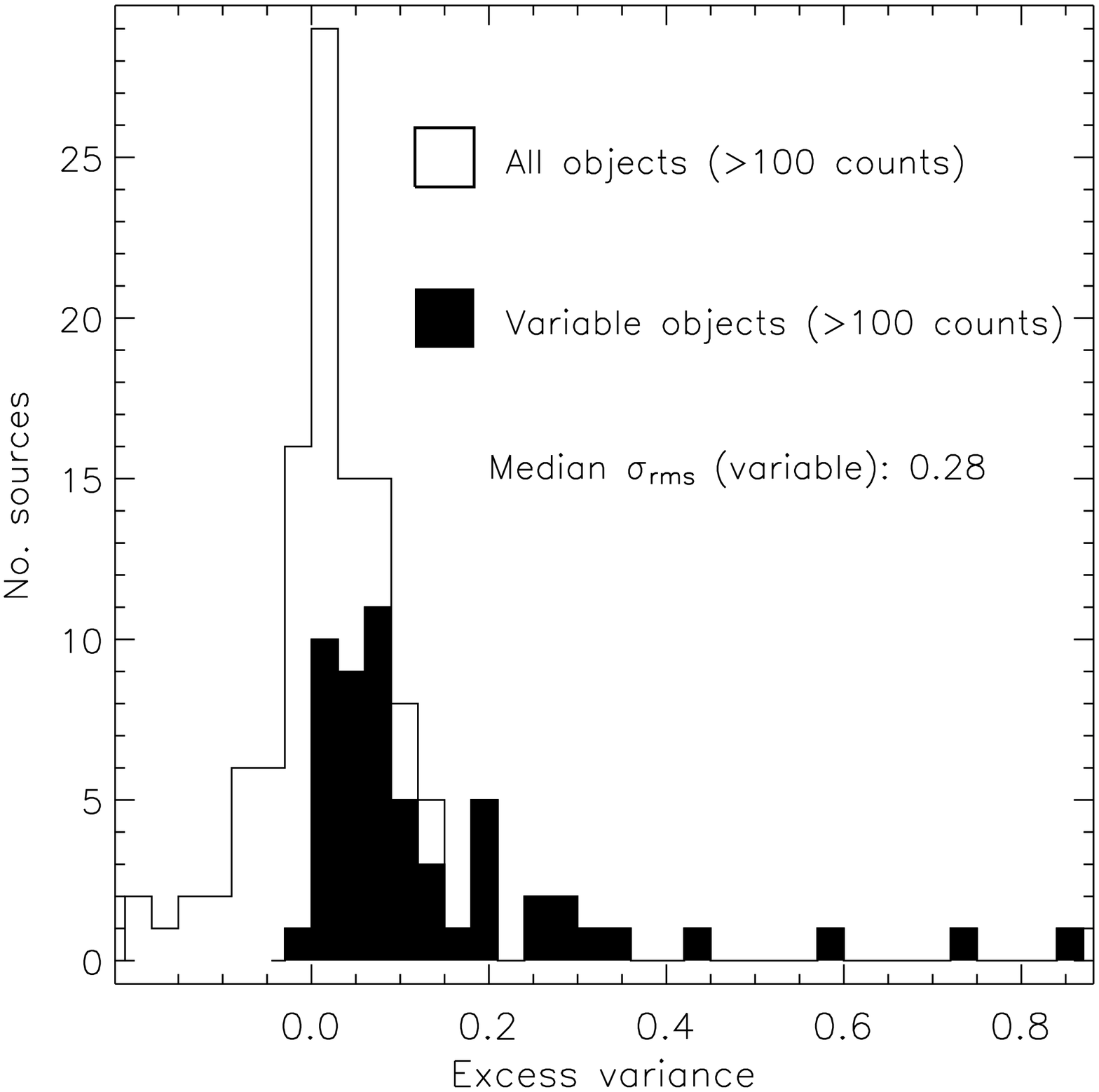}{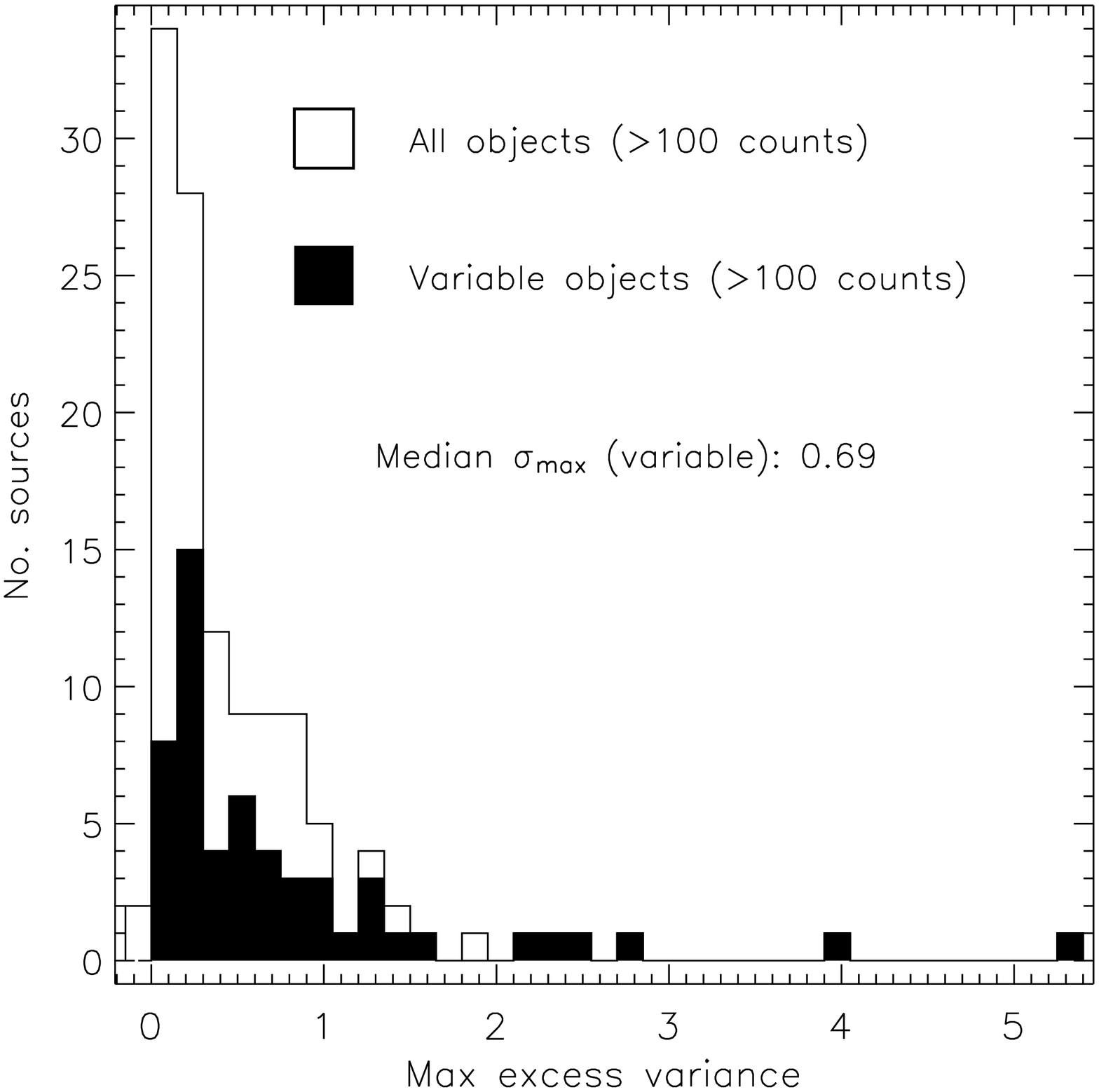}
\caption{Distribution of the excess variance ($\sigma_{rms}^2$) and max.excess variance  ($\sigma_{max}^2$) in the CDFS sources. The median flux variations (median $\sigma_{rms}$ and  $\sigma_{max}$) are also shown.\label{excess_var_histo}}
\end{figure*}

The lightcurves of our X-ray sources are very heterogeneous: we find some sources which exhibit smooth flux declines and increases and others with completely random behaviors\footnote{Note that this is true also for sources with good statistics, where the random behavior cannot be attributed to statistical uncertainties.}.
The excess variance $\sigma^2_{rms}$ and the maximum excess variance $\sigma^2_{max}$
measure the (squared) fraction of the total flux which is variable in the observed bandpass, and these are useful quantities to characterize the source variability. In Figure \ref{excess_var_histo} we show the distribution of $\sigma^2_{rms}$ and $\sigma^2_{max}$ measured in the CDFS sources. The variable subsample spans a large range of $\sigma^2_{rms}$, ranging from a few \% up to $\sim 95\%$ of the total emission, with a median value of $\sigma_{rms}\sim 30\%$. The $\sigma^2_{max}$ distribution further shows that half of the variable subsample exhibits maximum flux variations $>90\%$, with peaks up to $>230\%$ ($\sigma^2_{max}=5.3$).
The distribution of both $\sigma^2_{rms}$ and $\sigma^2_{max}$ for variable sources represent the upper tail of the intrinsic distribution: there is in fact a large fraction of sources with low variable fluxes which can not be detected as variable with the present data due to the large statistical errors.


\subsubsection{Spectral variability}
\label{specvar_sec}
\begin{figure*}
\epsscale{2.0}
\plotone{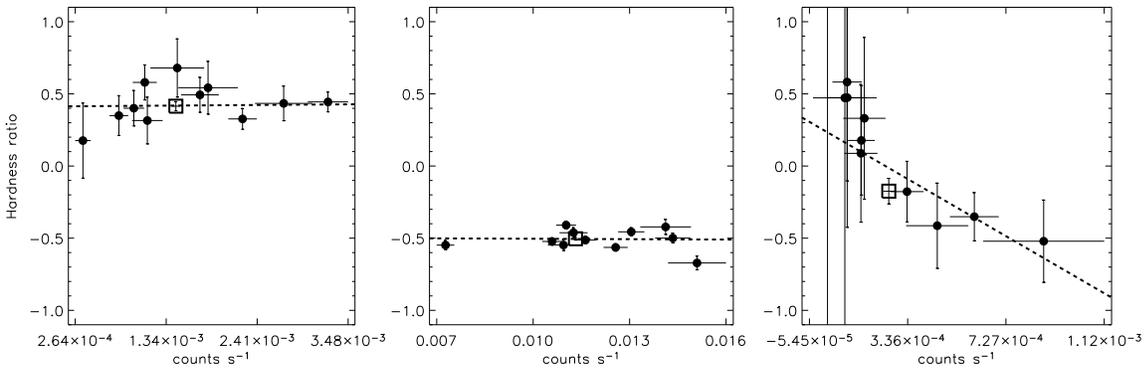}
\caption{Count rate vs hardness ratio for CDFS sources. The three panels show examples of sources with no spectral variability (left), spectral variability but no correlation with flux (center) and negative correlation with flux (right). Each dot represent a single exposure while the square shows the mean value obtained from the stacked observation. The best-fit line is also shown. \label{HRcorr}}
\end{figure*}

We studied the spectral variability of the sources by comparing the average HR of each source to the one measured in the individual exposures. Limiting the analysis to the 27 variable sources with $>500$ counts, where the statistics is high enough to allow an accurate HR determination in each bin, we find that only 4 sources exhibit spectral variability significant at the $2\sigma$ level (Figure \ref{HRcorr}). In 3 of them the variability is detected in just one exposure. 
We further looked for a correlation between total counts and HR, finding that 5 objects show significant correlation according to a rank correlation test: for all of them but one, the HR is anticorrelated with the total counts, meaning that as the source grows stronger, its spectrum gets softer (Figure \ref{HRcorr}, right panel). We point out that our sensitivity to HR changes is limited by the small statistics and varies depending on the total counts of the source. The average error on the HR in a single observation goes from $\sim 0.25$ at $n=100$ counts to 0.1 at $n=1000$, making power-law slope variations $\Delta\Gamma<0.2$ undetectable with the present data.

We conclude that $\sim 30\%$ of the sources with $>500$ counts show some sign of spectral variability, with spectral changes $\Delta\Gamma>0.2$. Furthermore, while we observe the presence of a softening of the spectrum with flux strength in $\sim 15\%$ of the sources, we find no convincing evidence of the existence of objects which show a spectral hardening. Note that we find no significant correlation with redshift, indicating that the lack of spectral variability in the majority of the sources is not due to the harder band sampled at higher $z$.

\subsubsection{Variability timescales}
\begin{figure*}
\epsscale{1.8}
\plotone{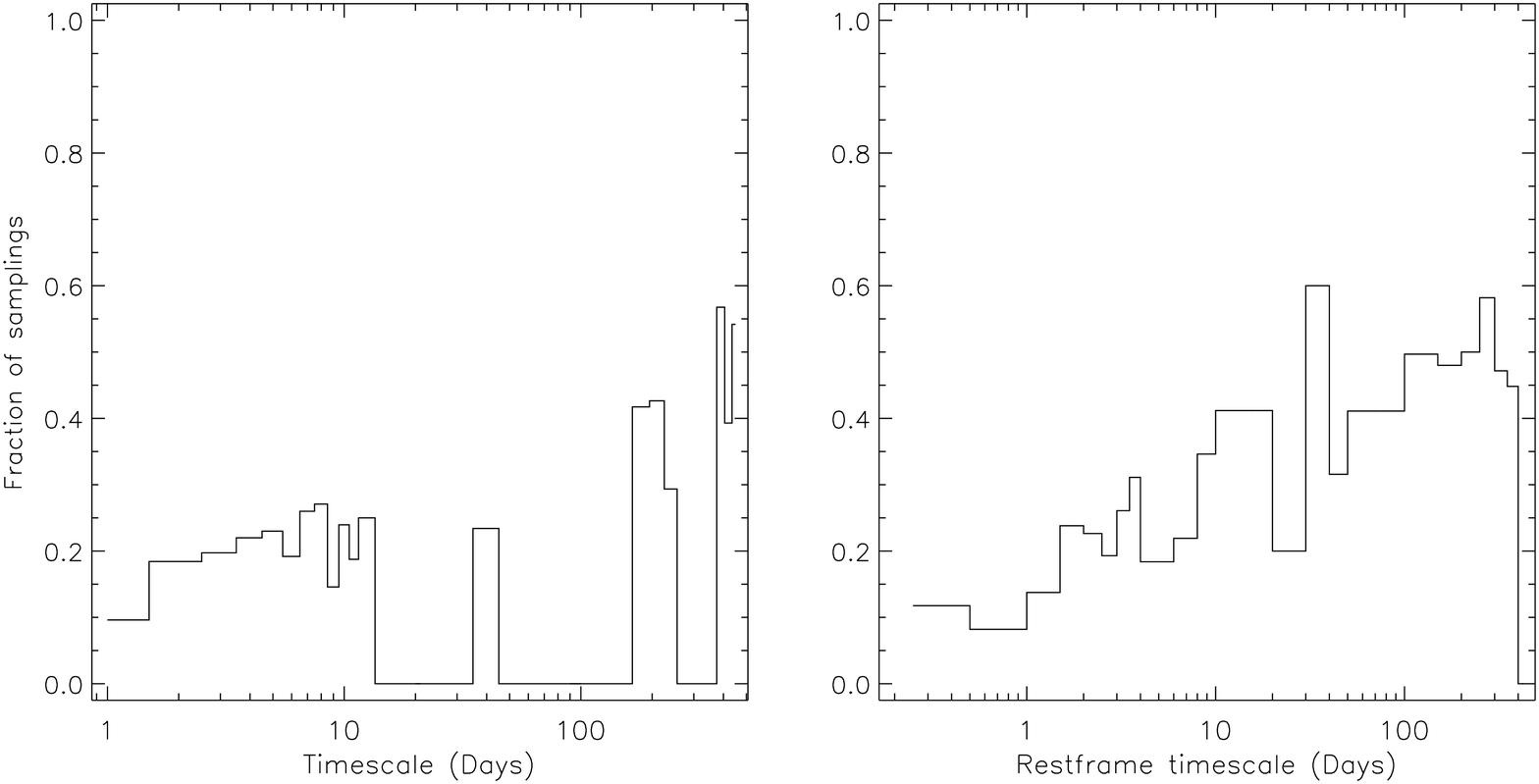}
\caption{{\bf Left panel}: Distribution of variability timescales in the observer rest-frame (0.5-7 keV band). The bin size is variable and the histogram has been normalized by the number of times each timescale has been sampled in the CDFS observations. The gaps are due to the irregular sampling of the CDFS field. {\bf Right panel}: timescale distribution in the source rest-frame for the subset of objects with available redshift. \label{timescales}}
\end{figure*}

To study the variability timescales of our sources we compared the difference $\Delta n=n_i-n_j$ in the net counts between each pair of bins $i,j$ where $i\in\{0,\cdots ,10\}$ and $j\geq i$,  with the one predicted by the simulations for a source with constant flux. We assumed that a source is significantly variable on a given timescale $\Delta t=|t_i-t_j|$ if the probability of observing a smaller $\Delta n$ is $>95\%$. 
The distribution of the measured timescales is shown in Figure \ref{timescales} (left panel). Each bin has been normalized to the number of times the corresponding timescale has been sampled in all sources to account for the irregular distribution of the X-ray observations and the missing data in the lightcurves of sources falling near the edge of the FOV. The gaps in the histogram represent timescales which have not been sampled by our data. In practice the histogram represents the probability of detecting variability on a given timescale $\Delta t$.

In the right panel we show the distribution of timescales $\Delta t_{rest}=\Delta t\cdot (1+z)^{-1}$ in the source rest-frame (i.e. correcting for time dilation) for the subsample with available photometric redshift. Due to the redshift distribution of the sources, this time we are able to sample all timescales ranging from less than one day up to a year, almost without any gap. We find evidence of variability on all timescales; however the figure shows that long-term variability prevails and that it is $\sim 5$ times more likely to observe variability on timescales $>100$ days than on less than a day. We stress that this difference is not due to a different sampling of short and long timescales since the histograms are normalized, nor to the different S/N ratio of the individual observations, since the average length of the December 2000 exposures (used to sample the short term variability; see Table \ref{obstab}) is comparable to the average of all exposures.

\begin{figure*}
\epsscale{1.0}
\plotone{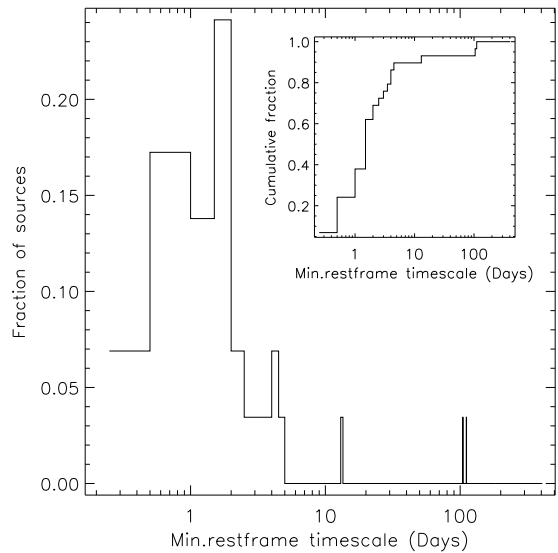}
\caption{ Distribution of the minimum restframe (right panel) timescales in the 0.5-7 keV band. The cumulative fraction is shown in the inner panel.\label{min_timescales}}
\end{figure*}

In Figure \ref{min_timescales} we present the distribution of the minimum rest-frame timescale observed in each lightcurve. The plot shows that 70\% of the variable sources possess variability on timescales smaller than 2 days. This means that even though long-term variability is dominant in our sample, in the sense of observing a source at two different times and finding a higher probability of detecting variability on long timescales than on short ones, short-term variability is still found in the majority of the sources.

\subsection{Luminosity-variability anticorrelation}
\label{Lxvar_sec}
\begin{figure*}
\epsscale{1.0}
\plotone{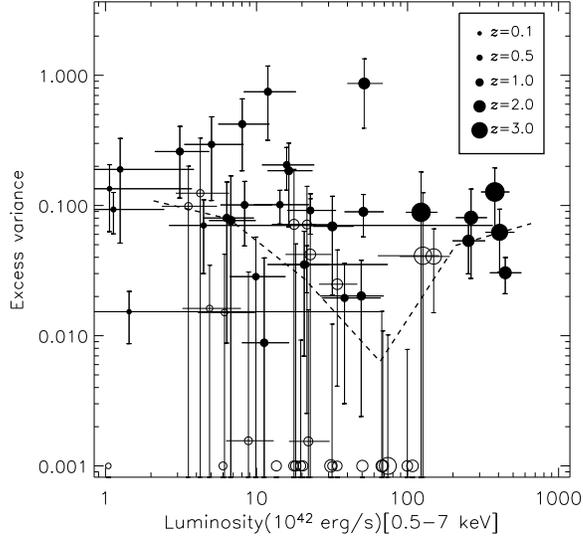}
\caption{$L_X$ vs excess variance. Filled (empty) circles represent variable (non-variable) sources and their size is coded according to redshift. Horizontal error bars reflect uncertainties on both the source flux and the photometric redshift. The dashed line shows the biweight mean of the excess variance calculated in logarithmic luminosity bins. Sources with negative excess variance are plotted at $\sigma_{rms}=10^{-3}$. \label{Lxvar}}
\end{figure*}

It is known that the variability amplitude of AGNs is anti-correlated with luminosity, in the sense that brighter sources exhibit smaller variable flux fractions than fainter ones \citep[e.g.][]{Barr86,Law93,Nandra97}. In Figure \ref{Lxvar} we plot the luminosity of the CDFS sources with available redshifts versus their excess variance. The luminosities are calculated in the 0.5-7 keV band and k-corrected assuming a power law spectrum with $\Gamma=1.8$. In this figure we see little correlation between the two quantities, as expected given the large errors both on the excess variance, due to the small statistics, and on the luminosities, reflecting the uncertainties in the photometric redshifts. To enhance any underlying $L_X$-variability correlation we divided the luminosity range in logarithmic bins and calculated the biweight mean (which is less sensitive to outliers than the regular mean) of the excess variance in each of them. We can see that in the $10^{42}<L_X<10^{44}$ range the average values show anticorrelation of the variable flux fraction with luminosity. However the brightest sources, at high redshifts, have large excess variance, inconsistent with extrapolating the trend of lower $z$ sources to high luminosities.
It may be argued that the use of a fixed energy band does not sample the same spectral range for sources at different $z$. In principle we do not expect that the use of different energy bands significantly affects the conclusions since the majority of our sources do not show dependence of variability on the energy range (see $\S$ \ref{specvar_sec}). However, to further test this, we extracted luminosities and variabilities in the 2-7 keV rest-frame band finding that, while the scatter of the sources is increased due to the smaller statistics (we are using less than half of the detected photons), the high $z$ sources still have systematically higher variability inconsistent with low $z$ sources.

\begin{figure*}
\epsscale{1.0}
\plotone{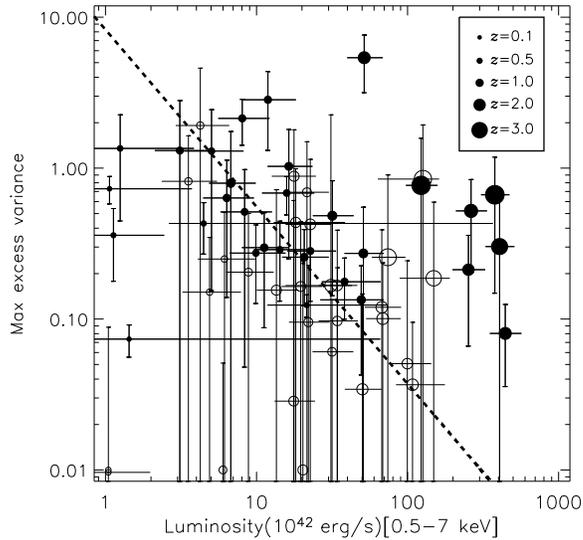}
\caption{$L_X$ vs max.excess variance. Symbols have the same meaning as in Figure \ref{Lxvar}. The dashed line shows the best fit to sources with $0.5<z<1.3$. \label{Lxmaxvar}}
\end{figure*}

Using the maximum excess variance $\sigma_{max}$ (Figure \ref{Lxmaxvar}) the correlation for sources in the redshift range $0.5<z<1.3$ is more evident. Note that this is the redshift range where we find the majority of our sources and where we are able to sample a large range of luminosities (Figure \ref{L-z}). This is due to the smaller uncertainties associated with $\sigma_{max}$ and, to a lesser extent, to the fact that the use of $\sigma_{max}$ tends to reduce the scatter due to the random sampling of our sources. A least squares bisector fit (allowing for uncertainties in both coordinates, \citealp{Isobe90}) yields a power-law slope of $\sim-1.17\pm 0.14$ ($-1.31\pm 0.23$ using $\sigma_{rms}$ instead of $\sigma_{max}$). Again high redshift sources are significantly more variable, while sources at $z<0.5$ seem to show the opposite trend, lying preferentially in the left region of the plot. We point out that no correction for time dilation has been applied to the data since it would require to make a somewhat arbitrary assumption about the Power Density Spectrum (PDS) of our sources (which we cannot reliably measure with the present data). However note that assuming a power-law PDS with a slope $\sim 1.5$, as seen in local AGNs, would further enhance the variability of high redshift sources. 

\subsection{Dependence of variable fractions on redshift}
\label{varfrac_z_sec}
\begin{figure*}
\epsscale{1.0}
\plotone{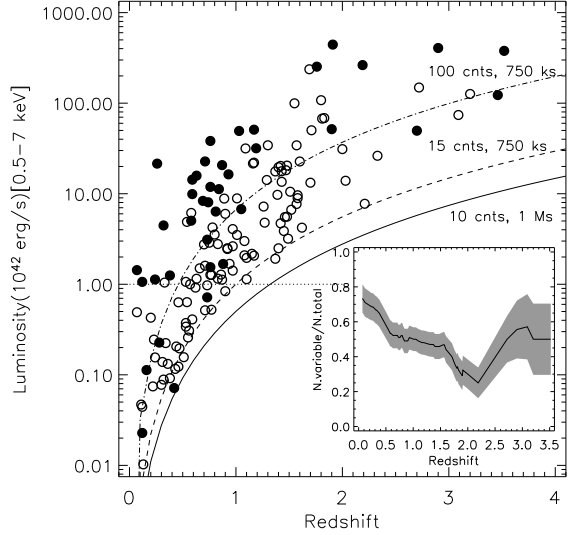}
\caption{Luminosity vs redshift. Variable and non-variable sources are
represented respectively by filled and empty circles. The horizontal dotted line marks the L$_x=10^{42}$ erg s$^{-1}$ limit. The curves show the minimum CDFS detection limit (continuous line), the average limit (dashed line) and the 100 counts limit (dot-dashed line). {\bf Inner panel:} Fraction of variable sources vs redshift. The fraction is calculated using a running average with a bin size of $\Delta z=2$ except at the edges of the data range where $\Delta z$ is truncated. The shaded region shows the statistical uncertainty; it does not include the dependence on the average source flux, which is found to be negligible. \label{L-z}}
\end{figure*}

In Figure \ref{L-z} we plot the luminosity of CDFS sources as a function of photometric redshift. To study the dependence of the fraction of variable sources $N_{var}/N_{Tot}$ on redshift, we measued the running average value of $N_{var}/N_{Tot}$ using a bin width\footnote{At the edges of the interval covered by our data the bin size is truncated on one side, thus reaching an effective width of $\Delta z=1$ for $z=0$ and $z=3.5$} $\Delta z=2$ (Figure \ref{L-z}, inner panel), limiting the analysis to objects with $>100$ counts and $L_X>10^{42}$ erg s$^{-1}$.
The large bin width minimizes uncertainties due to variations in the source flux. The median source counts ($\simeq 300$) are constant  with redshift within $\sim 15\%$ and thus their distribution does not affect significantly our variable/non-variable classification (see Figure \ref{varfrac}). 

The plot shows that the fraction of variable sources decreases from $z=0$ to $z\sim 2$ and then tends to increase again at higher redshifts. We find that this trend is significant at the $\sim 2.5\sigma$ level. We checked for energy band and timescale selection effects. Using the rest-frame 2-7 keV band we confirmed the decreasing trend at the $2\sigma$ level (note that the lower significance level is mainly due to the smaller number of sources in the sample with $>100$ counts). Time dilation produces a difference in the rest-frame timescales between the medium and high redshift bins of a factor $\simeq 3$ which, according to Figure \ref{timescales} would affect the variable fractions to $<10\%$. 

\begin{figure*}
\epsscale{1.0}
\plotone{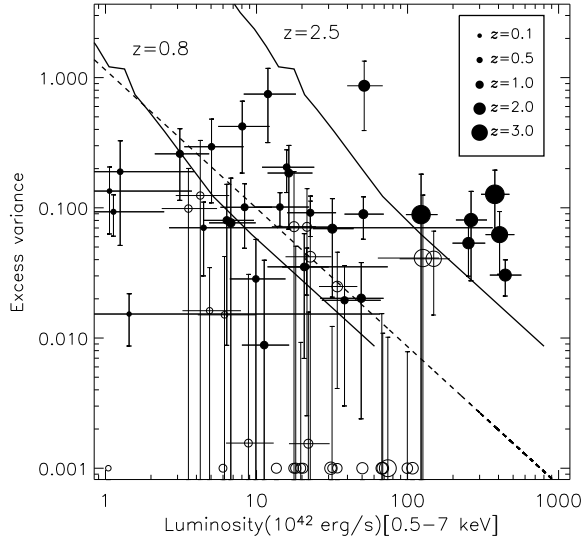}
\caption{$L_X$ vs excess variance. Symbols have the same meaning of Fig.\ref{Lxvar}. The dashed line represents the best fit to sources with $0.5<z<1.3$. The continuous lines show the average variability detection limit from Figure \ref{var_vs_cnts} for sources at $z=0.8$ and $z=2.5$. At high redshift the higher variability detection limit is balanced by the increase in excess variance of distant sources. \label{Lxvar_limits}}
\end{figure*}

We further note that the value of the $N_{var}/N_{Tot}$ ratio is scarcely affected by which luminosity range is sampled at any given redshifts because, while more luminous sources are less variable based on the $L_X$-variability relation, they also have larger fluxes making variability more easily detectable. Figure \ref{Lxvar_limits} shows that these two effects closely balance each other, since the average variability detection limit for sources at a given redshift has the same slope as the $L_x-\sigma^2_{rms}$ correlation. On the other hand if all AGN obey the same (in terms of slope and normalization) $L_X-\sigma^2_{rms}$ correlation of local AGNs, we wouldn't expect to detect any variable source at high redshift where we sample only the brightest objects. The fact that we do detect variability in distant AGNs is due to the increased $\sigma^2_{rms}$ of these sources which balances the higher variability detection limit (Figure \ref{Lxvar_limits}). Thus, while an increase $N_{var}/N_{Tot}$ ratio at high redshift is explained by the the high-luminosity sources which are more variable than their low-z counterparts, the decreasing trend for $z<2$ is harder to explain. We discuss some tentative interpretations of this result in $\S$ \ref{varfrac_disc}.

\section{Discussion}
\subsection{Prevalence of variability}
The analysis of the X-ray sources detected in the CDFS reveals that ($\S$ \ref{varfrac_sec}) 20\% of the sample is variable in the 0.5-7 keV band. However, excluding sources with too few counts to measure variability ($<100$ counts), this fraction becomes $45\%$ and increases with total flux up to 94\% at 800 counts, implying that we detect variability whenever we have enough photon statistics to measure it.
This is consistent with the analysis performed by \citet{Bauer02,Bauer03} in the {\it Chandra} Deep Field North, where the authors detect a fraction of variable sources as high as $90\%$ when adequate statistics is available.

This result suggests that the vast majority of AGNs possess X-ray variability on timescales ranging from days to months. However we have also shown ($\S$ \ref{HRvar_sec}) that the flux of the sources is correlated with their hardness ratio, in the sense that the brightest sources tend to be softer than the fainter ones. We studied the dependence of the variable fraction on counts for the hard and soft subsamples separately and we found the same increase in the variable fraction with counts, eventually reaching 100\%, even though the significance of the results is lower than for the whole sample due to the smaller number of sources. We conclude that that both hard and soft populations are highly variable, even though in harder sources our ability to detect flux variations is reduced due to a combination of smaller intrinsic fluctuations and a lower average flux.

As discussed in $\S$ \ref{char_sec}, our sources are very heterogeneous in terms of variability: most lightcurves are characterized by random fluctuations, some also show well defined trends with flux slowly increasing or decreasing over several months. This complex behavior is represented in the histograms in Figure \ref{timescales} which show variability at all timescales from a fraction of a day, up to a year. Long term variability dominates our sample: detecting variability on timescales longer than a month is twice as likely than on timescales smaller than 10 days. This is in qualitative agreement with the power-law shape of the PDS characterizing the variability of local AGNs, which implies larger variability amplitudes on long timescales and thus a larger chance of detecting flux variations. On the other hand we detect short-term variability ($<2$ days) in the majority of our sources, meaning that part of the flux variations are produced on spatial scales $<2\times 10^{-3}$ pc, corresponding to the inner part of the accretion disk and and of the Broad Emission Line region. This suggests that whatever process is responsible for the variability must account for a wide range of timescales or that several different processes are contributing at the same time.

We do not find any significant ($\Delta \Gamma>0.2$) spectral variability in 70\% of the sample ($\S$ \ref{specvar_sec}). In half of the remaining sources the HR appears anti-correlated with the flux, similar to the high-soft/low-hard behavior of galactic sources \citep{Klis95}. This lack of correlation between variability and spectrum in most sources argues against variable obscuration as the main mechanism responsible for X-ray variability, even though small column density changes would not be detectable in the present data. The presence of sources in which spectral variations are indeed observed but appear uncorrelated with flux, indicates that in these objects the spectral changes are not triggered by X-ray luminosity fluctuations.

\subsection{Variability dependence on absorption}
\label{abs_disc}
In $\S$ \ref{HRvar_sec} we showed that the variable and non-variable sources represent two distinct populations in terms of their absorption. The observation that variable sources are significantly softer, on average, than non-variable ones can be explained by invoking different column densities of absorbing material, which affects mostly photons below 2 keV, making sources both fainter and harder (Figure \ref{HR_vs_cnts} and \ref{abs}). 
This separation between variable and non-variable sources according to intrinsic absorption mimics closely the one between Type I and Type II AGNs shown in Figure 9 of \citet{Tozzi01}.
Even though it is difficult to separate the actual dependence of variability on  absorption from the spurious effect introduced by the lower statistics in fainter sources (which makes variability harder to detect) our tests suggest, at the $2\sigma$ confidence level, that there is an intrinsic difference in the variability of the {\it absorbed} and {\it unabsorbed} populations.

Previous works also found a correlation between the spectral properties of the sources and their variability, in the sense that the more variable sources are those with steeper spectra \cite[e.g.][]{GML93,Koenig97,Turner99,Grupe01}. The origin of such correlation is still far from clear: both parameters (variability and spectral shape) may be related to BH mass and accretion rate, which would induce different degrees of variability while, at the same time, creating different conditions in the accretion disk and its corona (optical depth, temperature) explaining the different spectra \citep[see for instance][]{Turner99}. Reprocessing of the original radiation has also been invoked \citep[e.g.][]{GML93,Koenig97}, since high energy photons would be produced through a larger number of collision which also tend to smear out the variability.

We note that these models have been invoked to account for observations of short timescales variability and moderate spectral changes ($\Delta \Gamma\lesssim 1$) in samples of Seyfert 1 galaxies, i.e. where we are observing more or less directly the central engine. Our data, instead, span a large range in HR, implying variations in the photon index up to $\sim 2$ and large column densities. The hardest sources in our sample are in fact consistent with column densities found in type II AGNs. An alternative explanation of the correlation between variability and absorption in the CDFS sources is that the X-ray observed emission is due to two main components: a highly variable one which originates near the central engine and a reflected one, produced by material distributed over a large region which dilutes the intrinsic variability of the primary component. When the central engine is observed directly, i.e. through low column densities, the first component dominates the total flux and thus the variability. In high $N_H$ sources, instead, the primary component is highly absorbed. Then, if the reflecting material is distributed at large distances from the central BH or in a geometrical configuration such that it is relatively unaffected by the obscuring material (e.g. the molecular torus in the classic unification scheme), the scattered component will represent an increasingly larger fraction of the observed emission thus accounting for the lower variability. 
This scenario is supported by the recent detection of a soft component in the XMM spectra of several  distant type-2 QSO in the CDFS field \citep{Streb03}, which is well fitted by a scattering model.

\subsection{Dependence of the variability amplitude on X-ray luminosity and redshift}
\label{L_x_var_sec}
An anti-correlation between luminosity and variability of AGNs in the X-ray band was originally suggested by \cite{Barr86} and later confirmed by several authors \citep[e.g.][]{Law93,GML93,Nandra97}. Other works, however, pointed out that the luminosity-variability correlation is affected by significant scatter and may depend on the AGN properties (line width, spectral shape etc.; \citealp{Nandra97,Turner99,Fio98}).

Given these premises, it is not surprising that in our data it is difficult to clearly identify a correlation between these two quantities, since we lack detailed information on the properties of CDFS sources and are thus likely to be including many different classes of objects in our sample. The problem is further enhanced by the low statistics and by the sparse sampling of the lightcurves. Nevertheless we have shown in $\S$ \ref{Lxvar_sec} that our observations are consistent with the presence of such anti-correlation, particularly when we use the maximum excess variance which is less affected by statistical errors and partly removes the effect of the random sampling.

A more quantitative comparison with previous works is difficult because the large uncertainties in our data do not allow to put precise constraints on the slope of the $L_X-$variability relation. Restricting our analysis to sources in the range $0.5<z<1.3$ we obtain a power-law slope of $-1.31\pm 0.23$ (or $-1.17\pm 0.14$ using $\sigma^2_{max}$), steeper than the average slope $\simeq -0.6$ usually found in literature for local AGNs (see for instance \citealp{Law93,GML93,Nandra97}). Note however that these studies sampled nearby ($z<0.1$) AGNs on short timescales (less than one day). At higher redshift the situation is less clear: \citet{Almaini00} found a slope of $-1.5\pm 0.2$ analysing   a deep {\it ROSAT} sample of AGNs at $z<0.5$ on timescales of days to weeks. On the other hand the enlarged {\it ROSAT} sample studied by \citet{Manners02} yields a value of $0.54\pm 0.10$, closer to the local one.
The heterogeneous composition of moderate and high redshift samples, for which accurate spectroscopic classification are not available,  may explain the steeper slope: \citet{Turner99}, for instance, show that the inclusion of narrow line Seyfert galaxies in their sample introduces large scatter, possibly increasing the slope. A better selected sample is required to properly address this issue.

A more interesting result is our finding that high redshift sources behave differently, in terms of variability, than their low redshift counterparts ($\S$ \ref{Lxvar_sec}). Sources with $z\gtrsim 1.3$ have large excess variances which are inconsistent with those expected from intermediate-$z$ ($0.5<z<1.3$) sources of comparable luminosities. This agrees with \citet{Almaini00} and \citet{Manners02}, which found that their most distant and luminous sources do not obey the same correlation of nearby ones, being more variable. Both redshift evolution and luminosity can be invoked as the parameters producing the "upturn" at high $L_x$, given the degeneracy between the two quantities in flux limited samples. Unfortunately the luminosity ranges sampled by intermediate and high $z$ sources in our data (as in those of Almaini and Manners) scarcely overlap, and thus do not permit determining if the $L_X-$variability relation for  intermediate-$z$ sources extends to higher luminosities. Nevertheless the faintest source ($L_X\simeq 10^{43}$) in our high-$z$ subsample is highly variable ($\sigma^2_{rms}\sim 1$, see Figure \ref{Lxmaxvar}), suggesting that a correlation between $L_X$ and $\sigma^2$, with similar slope to the one found for intermediate objects, exists in distant sources. Furthermore sources with $z<0.5$ tend to lie on the lower-left side of the diagram: unless we assume that there is also a "downturn" of the $L_X-$variability relation at the lowest luminosities, this supports the idea that evolution shifts the whole correlation toward lower variabilities and/or luminosities. Comparing our data with \cite{Almaini00} and \citet{Manners02} we also note that their "upturn" occurs at much higher luminosities than in our data. Thus the interpretation that it may reflect a maximum in the black hole size distribution \citep{Almaini00} does not hold for our sample. 
The most likely scenario is that the same anti-correlation is present up to at least $z\sim 3$ but that it evolves with redshit, in the sense that objects with comparable luminosities were more variable in the past. The "upturn" is produced by selection effects which prevent us from observing low luminosity AGNs at high redshift.

Possible interpretations of this result have been explored by Manners and collaborators (2002). They include an intrinsic change in AGN luminosity triggered by higher accretion efficiency or a change in variability either due to a smaller X-ray emitting region or to fainter X-ray flares. Our data further support these interpretations. In fact we can rule out the possibility that the observed increase in variability at high $z$ is due to bandpass effects, because we are using a broader spectral band than the ROSAT one used by \citet{Manners02} and because we detected the same trend using a fixed rest-frame band ($\S$ \ref{Lxvar_sec}). The possibility that larger variability is due to obscuration provoked by intervening obscuring material across the line-of-sight seems also unlikely, since no correlation between flux and hardness ratio is detected in the majority of our sources ($\S$ \ref{specvar_sec}). Finally it is possible that the emergence of a population of highly variable AGN at high redshift, such as narrow-line Seyfert 1, may increase the variability in the high redshift subsample. However the preliminary inspection of the optical spectra of CDFS sources, obtained by Szokoly et al. (2004, in preparation), suggests that our most distant X-ray sources include both narrow and broad emission line AGNs whose relative abundance does not differ from the one found at lower redshifts.


\subsection{Variable fraction evolution}
\label{varfrac_disc}
In $\S$ \ref{varfrac_z_sec} we showed that the fraction of variable sources decreases with redshift for $0.5< z\lesssim 2$. While the significance of the result ($<3\sigma$) does not allow to exclude that this decrease is due to statistical uncertainties, we argued that it is not due to systematic effects due to either flux, timescale or bandpass selection biases. On the other hand luminosity selection effects may produce the observed trend since distant bright sources are less variable than their faint nearby counterparts. This interpretation however  requires that all sources $z<2$ follow approximately the same anticorrelation, and any significant increase in variability occurs only at larger redshifts. This possibility cannot be excluded based on the  $L_X-\sigma^2_{rms}$ correlation shown in Figure \ref{Lxvar} and would be consistent with \cite{Manners02}, who also found larger variability amplitude only for $z\gtrsim 2$. 

Alternatively we note that \cite{Gilli03} detected the presence of a cosmic structure at $z\sim 0.7$ in the CDFS. This is also the redshift where most of our variable sources are located. If environmental properties trigger AGN variability, it is possible that the large variable fraction at low $z$ is due to the presence of such a structure.
Finally, if variability is correlated with intrinsic absorption and the variable/non-variable separation reflects the Type I/Type II classification, as discussed in $\S$ \ref{abs_disc}, the decrease in variable fraction may reflect the evolution of the Type I/II ratio. In fact AGN population synthesis models trying to explain the observed X-ray background, predict a decrease of the Type I/II ratio by a factor $\geq 2$ between $z=0$ and $z\simeq 1.3$ \citep{Gilli01}.

Future work using spectroscopic redshifts and an extended sample to minimize the statistical uncertainties, possibly including the {\it Chandra} Deep Field North, may be decisive in better constraining the evolution of AGN variability at high redshift.

\section{Conclusions}
We exploited the deep (1 Ms) and high-resolution data of the {\it Chandra} Deep Field South to study the temporal variability of AGNs on timescales ranging from a fraction of a day up to one year. While similar studies have been performed on samples of nearby AGNs, our sample primarily contains moderate luminosity AGNs out to $z\simeq 3.5$. 

We detect significant variability in $>50\%$ of the AGN population in the 0.5-7 keV band and suggest that as many as $90\%$ of all AGN in our sample may be variable. The variability exists on all timescales that we can observe, but our sources are twice as variable on timescales of $>100$ days than within a few ($<10$) days. All sources show some variability on the shortest timescales indicating that part of the flux variations are produced on spatial scales $<2\times 10^{-3}$ pc, corresponding to the inner part of the accretion disk and of the Broad Emission Line region. 

We also find that unabsorbed sources ($\log N_H<22.5$ cm$^{-2}$) appear intrinsically more variable than absorbed ones at the $2\sigma$ significance level, suggesting that the lack of variability in hard, absorbed sources is due to an increased contribution of reflected X-rays to the total flux. On the other hand we do not find strong ($\Delta\Gamma >0.2$) spectral changes associated with the flux variations in the majority of our sources.

Our data are consistent with an anti-correlation between X-ray luminosity and variability, similar to that found in local AGNs. However, this relation seems to evolve with redshift, with X-ray sources having been more variable in the past. This trend may reflect an evolution in the accretion rate and/or size of the X-ray emitting region with look-back time.

Finally we find a possible decrease in the fraction of variable sources with redshift for $0.5\lesssim z<2$. This result could be explained by assuming that AGN variability has not evolved since $z\sim 2$. Alternatively, environmental effects triggering X-ray variability or an evolution in the Type I/Type II AGN ratio could produce the observed trend. Future work using spectrosopic redshift and an enlarged sample may be able to address this issue.

\clearpage
\appendix
\centerline{\bf\large Appendix}
\section{X-ray sources simulations}
\label{app_simul}
To simulate the lightcurve of a source in the CDFS, in the assumption that the X-ray flux remains constant in time, we used the following procedure:
\begin{enumerate}
\item we took the average net source count-rate, measured from the stacked observation as described in $\S$ \ref{ltc_sec}, and rescaled it to the actual exposure (including the exposure time and the position-dependent sensitivity correction) of the real source in each individual observation, obtaining the average expected counts $\overline n_s$ within the source region for a non-variable source;
\item the average background counts expected within the same region were derived in a similar way, using the average background count-rate $\overline n_b$ measured in the background region, rescaled to the source to background area ratio $A_s/A_b$;
\item the simulated counts within the source region are then given by $C_{tot}=C_s+C_b$, where the contribution of the source and the background, $C_s$ and $C_b$, were randomly extracted from a Poisson distribution with mean value $\overline n_s$ and $(\overline n_b*A_s/A_b)$ respectively;
\item the counts expected from the background region were accordingly sampled from a Poissonian distribution with mean value $\overline n_b$.
\end{enumerate}
This simulated observation of the source and the background for each exposure includes the dependence on the position of the source on the detector, the actual exposure time and obeys the correct statistics.

\section{Sources of uncertainty in the variability estimates}
\label{uncertainty_app}
{\it QE degradation}: it is known that the \chandra ACIS quantum efficiency (QE) has undergone a continuous degradation since launch in 1999, due to molecular contamination building up on the cold optical blocking filter, and/or the CCD chips. This degradation is most severe at low energies below 1 keV, producing a decrease in the QE at 670 eV of about 10\% per year\footnote{More details at http://cxc.harvard.edu/cal/Acis/Cal\_prods/qeDeg/}. This effect may introduce spurious variability in our sources, since the observations span a large time interval. To estimate the influence of the QE degradations we used the ACISABS absorption model of the {\it Sherpa} fitting software, included in the CIAO 2.3 package. We simulated the spectrum of an average CDFS source, composed of a power law with photon index $\Gamma=1.44$ absorbed by the galactic column density $N_H=8\times 10^{-19}$ cm$^{-2}$ \citep{Tozzi01}. We then applied to this model the additional absorption estimated by the ACISABS model for each individual observation, and we calculated the flux ratio between each observation and the last one (ObsID 2239). This factor estimates the decrease of the observed flux at each observing epoch and was used to normalize the net source counts to the last observation. Not surprisingly, the correction was found to be significant only for the first 4 exposures, since the following ones were acquired very close to each other (i.e. within 13 days, see Table \ref{obstab}). The correction factors were estimated to be 6\% for ObsID 1431-0 and 1431-1, and 2\% for 441 and 582 in the 0.5-7 keV band, and slightly larger in the soft (0.5-2) keV band: respectively 9\% (3\%) in the first (second) two observations. The hard (2-7keV) band is unaffected by this problem, yielding corrections smaller than 1\%.

{\it Exposure correction}: the actual number of counts measured by \chandra for any CDFS source in a single observation depends on the source position on the detector due to the decrease of the mirror's effective area as a function of off-axis angle and QE variations of the ACIS CCDs. This difference is taken into account in our analysis through the exposure correction, normalizing the observed count rates to those expected at the aimpoint ($\S$ \ref{ltc_sec}). The exposure maps themselves however, depend on the spectral energy distribution (SED) of the source. Since for most sources there are not enough counts to obtain a reliable SED, we used the average hard exposure maps centered at 4.5 keV from \citet{G02}. If a source has a very soft spectrum however, this may introduce spurious variations in the exposure-corrected counts. We estimated the influence of the exposure correction comparing the hard exposure maps to those produced for the soft band and centered at 1.5 keV. We found that the maximum difference between soft and hard exposure maps is $\leq 8\%$ except in ObsID 1431-0 and 1431-1 where it is $\sim 10\%$. The CDFS observations have approximately the same aimpoint so that the source position on the detector changes mostly along the azimuthal direction, with small variations in the distance from the aimpoint. Since the exposure maps are largely azimuthally symmetric as well, these uncertainties due to the exposure corrections must be considered upper limits, with most sources affected to a much lesser extent. 

{\it PSF variations}: another concern in studying temporal variability is the influence of using a fixed extraction radius, while the PSF size varies between observations due to the different source position on the detector. The influence of a varying PSF, however, is minimized by the fact that the main parameter affecting the PSF size is the radial distance from the aimpoint\footnote{See the \chandra Proposers' Observatory Guide: http://cxc.harvard.edu/proposer/POG/html/MPOG.html}. Since the CDFS observations have only small offsets ($< 20"$) the changes in the PSF size are minimal. We quantified this uncertainty factor by extracting encircled energy profiles for a few bright off-axis sources in our field: we found that the use of the average 95\%  encircled energy radius ($\S$ \ref{ltc_sec}) produces count uncertainties smaller than 2\% between different exposures.

{\it Sources on edges}: we noted in $\S$ \ref{ltc_sec} that, in order to maximize the number of bins for each source, we retained in our lightcurves observations where the source falls in part outside the FOV. The criteria for inclusion were chosen to minimize the effect of the missing area on the count extraction. The requirement to have at least $50\%$ of the original background area allows a reliable background estimate albeit with reduced accuracy; this is properly taken into account in the error estimate. When the source region falls partly outside the FOV instead, we estimated that the 10\% upper limit on the missing area yields an underestimate of the source counts $<2\%$, since the affected region lies on the PSF wings.  

\vspace{2cm}
{\bf \large Acknowledgements:} The authors whish to thank Paolo Tozzi for several enlightening discussions. We also thank Franz E. Bauer for useful suggestions on the data analysis. We are grateful to the referee for numerous comments which improved the manuscript. We acknowledge financial support provided by NASA through GO grants GO08809.01-A and GO08119.02-A from the Space Telescope Science Institute, which is operated by AURA, Inc., under NASA contract NAS 5-26555. This research has made use of NASA's Astrophysics Data System Bibliographic Services.

\clearpage
\begin{deluxetable}{llrlll}
\tabletypesize{\scriptsize} \tablewidth{0pt} \tablecaption{\chandra 
observations of CDFS \label{obstab}} \tablehead{ \colhead{Obs. ID} &
\colhead{Obs. Date} & \colhead{Exp. Time (ks) \tablenotemark{a}} &
\colhead{Roll Angle} & \colhead{Aim Point} } 
\startdata 
1431-0 & 1999 Oct 15 & 24.983~~~ & 47.28 & 3 32 29.4 -27 48 21.8\\  
1431-1 & 1999 Nov 23 & 92.807~~~ & 47.28 & 3 32 29.4 -27 48 21.8\\ 
441 & 2000 May 27 & 55.727~~~ & 166.73 & 3 32 26.8 -27 48 17.4\\ 
582 & 2000 Jun 03 & 129.869~~~ & 162.93 &3 32 26.8 -27 48 16.4\\ 
2406 & 2000 Dec 10 & 29.564~~~ & 332.18 &3 32 28.4 -27 48 39.3\\ 
2405 & 2000 Dec 11 & 59.363~~~ & 331.81 &3 32 29.0 -27 48 46.4\\ 
2312 & 2000 Dec 13 & 123.212~~~ & 329.92 &3 32 28.4 -27 48 39.8\\ 
1672 & 2000 Dec 16 & 94.564~~~ & 326.90 &3 32 28.9 -27 48 47.5\\ 
2409 & 2000 Dec 19  & 68.719~~~ & 319.21 &3 32 28.2 -27 48 41.8\\ 
2313 & 2000 Dec 21 & 129.937~~~ & 319.21 &3 32 28.2 -27 48 41.9\\ 
2239 & 2000 Dec 23 & 130.250~~~ & 319.21 &3 32 28.2 -27 48 41.8\\ 
\enddata
\tablenotetext{a}{effective exposure time after cleaning bad aspect
interval and high background intervals (about 8800 s are lost due to
high background)} 
\end{deluxetable}


\clearpage

\clearpage

\clearpage

\clearpage

\clearpage

\end{document}